\documentclass[preprintnumbers,amsmath,amssymb,aps,pra,floatfix,lengthcheck,superscriptaddress]{revtex4-2}

\usepackage{bm}
\usepackage{graphicx}
\usepackage{color}
\usepackage{amsmath}
\usepackage{hyperref}
\usepackage{epstopdf}
\usepackage{upgreek}
\usepackage{mathptmx, textcomp}
\usepackage[latin1]{inputenc}
\usepackage[T1]{fontenc}
\usepackage{array, multirow}
\usepackage[table]{xcolor}
\usepackage{booktabs}
\usepackage{longtable}
\usepackage[normalem]{ulem}

\newcommand{\Ba}{Ba$^+$}
\newcommand{\BaRb}{BaRb$^+$}

\newcommand{\Rbplus}{Rb$^+$ }
\newcommand{\Rbmolplus}{Rb$_2^+$ }

\newcommand{\commentOut}[1]{}

\begin{document}

\title{Inelastic collision dynamics of a single cold ion immersed in a Bose-Einstein condensate}
\author{T. Dieterle}
\affiliation{5. Physikalisches Institut and Center for Integrated Quantum Science and Technology, Universit\"{a}t Stuttgart, Pfaffenwaldring 57, 70569 Stuttgart, Germany}
\author{M. Berngruber}
\affiliation{5. Physikalisches Institut and Center for Integrated Quantum Science and Technology, Universit\"{a}t Stuttgart, Pfaffenwaldring 57, 70569 Stuttgart, Germany}
\author{C. H\"olzl}
\affiliation{5. Physikalisches Institut and Center for Integrated Quantum Science and Technology, Universit\"{a}t Stuttgart, Pfaffenwaldring 57, 70569 Stuttgart, Germany}
\author{R. L\"{o}w}
\affiliation{5. Physikalisches Institut and Center for Integrated Quantum Science and Technology, Universit\"{a}t Stuttgart, Pfaffenwaldring 57, 70569 Stuttgart, Germany}
\author{K. Jachymski}
\affiliation{Faculty of Physics, University of Warsaw, Pasteura 5, 02-093 Warsaw, Poland}
\author{T. Pfau}
\affiliation{5. Physikalisches Institut and Center for Integrated Quantum Science and Technology, Universit\"{a}t Stuttgart, Pfaffenwaldring 57, 70569 Stuttgart, Germany}
\author{F. Meinert}
\affiliation{5. Physikalisches Institut and Center for Integrated Quantum Science and Technology, Universit\"{a}t Stuttgart, Pfaffenwaldring 57, 70569 Stuttgart, Germany}

\date{\today}

\begin{abstract}
We investigate inelastic collision dynamics of a single cold ion in a Bose-Einstein condensate. We observe rapid ion-atom-atom three-body recombination leading to formation of weakly bound molecular ions followed by secondary two-body molecule-atom collisions quenching the rovibrational states towards deeper binding energies. In contrast to previous studies exploiting hybrid ion traps, we work in an effectively field-free environment and generate a free low-energy ionic impurity directly from the atomic ensemble via Rydberg excitation and ionization. This allows us to implement an energy-resolved field-dissociation technique to trace the relaxation dynamics of the recombination products. Our observations are in good agreement with numerical simulations based on Langevin capture dynamics and provide complementary means to study stability and reaction dynamics of ionic impurities in ultracold quantum gases. 
\end{abstract}

\maketitle

Recent years have seen tremendous progress in studies of hybrid systems of trapped ions and ultracold atoms \cite{Tomza2019}. This platform currently offers means to study cold ion-atom collisions \cite{Zipkes2010,Meir2016,Dutta2018,Grier2009} and shows promising perspectives for accessing charge-neutral mixtures deep in the quantum regime \cite{Feldker2020,Schmidt2020,Kleinbach2018}. Ultimately, these advances are believed to allow for realizing strongly-coupled charge-neutral polaron systems \cite{Casteels2011,Astrakharchik2020}, for emulating solid-state Hamiltonians \cite{Bissbort2013}, or for studying novel few-body physics comprising mesoscopic molecular ions with large effective masses \cite{Cote2002, Schurer2017}. Besides reaching the celebrated quantum-scattering regime, a central point for these proposals is the stability of the ion in an ultracold and dense host gas. Similar to neutral atom systems \cite{Esry1999, Burt1997, Spethmann2012}, it was found that three-body recombination (TBR) may severely limit the lifetime of trapped ion-atom hybrids, specifically at high atomic densities. Large reaction rates for ion-atom-atom TBR, which are linked to the long-range charge-neutral $C_4$-interaction potential, have been observed \cite{Haerter2012,Kruekow2016,Haerter2014}.

While the rate constants for ion-atom-atom TBR in the semi-classical energy regime are comparatively well understood \cite{PerezRios2018}, much less is known about the subsequent kinetics of the reaction products, i.e. the kinetics of the produced molecular ions \cite{PerezRios2019,Jachymski2020}. In contrast to extensive studies of ultracold chemistry and inelastic collisions of deeply-bound trapped molecular ions \cite{Hudson2016, Sullivan2011,Rellegert2013, Hauser2015,Balakrishnan2016}, TBR produces very weakly bound molecular ions whose dynamics consequently occurs near the molecular dissociation threshold. Only very recently, evidence for secondary atom-molecule collisions has been reported for a system comprising a Paul trap for the ion and an optical dipole trap for the neutral atoms \cite{Mohammadi2020}.

\begin{figure}[!ht]
\centering
	\includegraphics[width=\columnwidth]{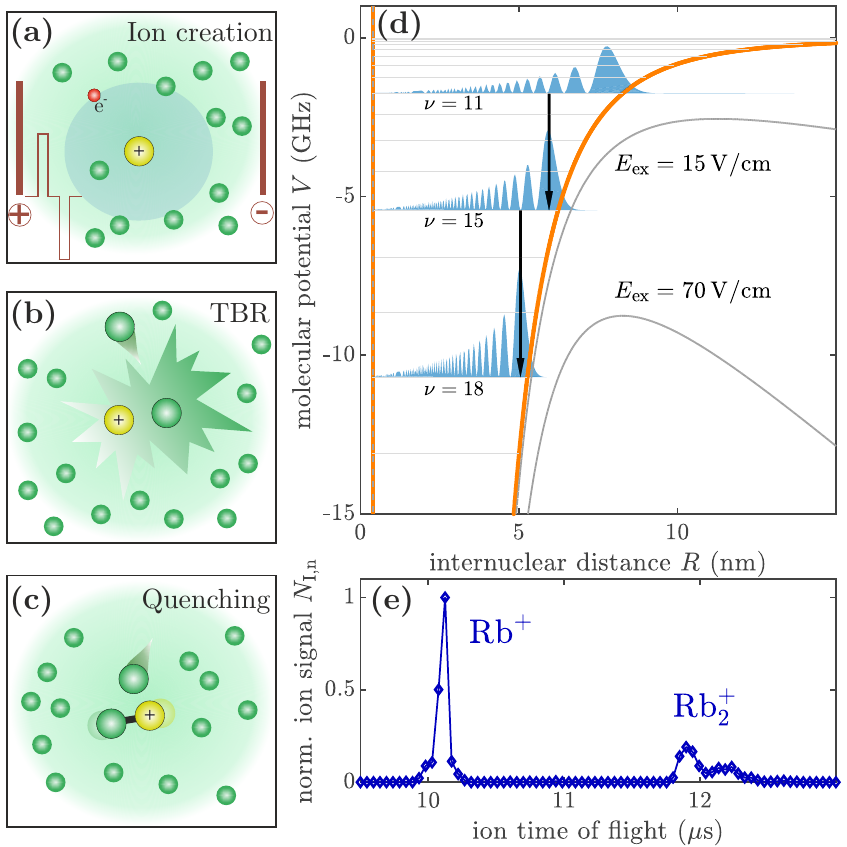}
	\caption{(a)-(c) Low-energy ion production and subsequent inelastic collision dynamics. (a) A single low-energy \Rbplus ionic impurity is implanted in a BEC via fast electric field ionization (inset indicates the two-pulse ionization sequence) of a precursor Rydberg atom. (b) Ion-atom-atom three-body recombination causes formation of a weakly bound \Rbmolplus molecular ion. (c) Secondary molecule-atom collisions lead to \Rbmolplus rovibrational quenching. (d) \Rbmolplus molecular potential $V$ as a function of internuclear distance $R$ near the dissociation limit. Binding energies of the most weakly bound vibrational states are denoted as horizontal lines together with three exemplary vibrational wavefunctions. Black arrows indicate vibrational quenching. An adjustable external electric field $E_{\rm{ex}}$ modifies the total potential (grey lines) and allows for energy-resolved dissociation of the molecular ion. (e) Time of flight distribution of ion counts on the MCP after $t = 21 \,\mu$s and for $E_{\rm{ex}}= 89\, \rm{V/cm}$ . \Rbplus ions (\Rbmolplus molecular ions) arrive around 10.1 $\mu$s (12 $\mu$s).}
	\label{Fig1}
\end{figure}

In this paper, we present studies of ion-atom-atom TBR and secondary collision-induced rovibrational relaxation of the weakly bound reaction products employing a platform complementary to the more conventional ion traps. Instead of holding the ion in a Paul trap, we implant a low-energy \Rbplus ion directly into a dense Bose-Einstein condensate (BEC) by exciting and ionizing a single Rydberg atom \cite{Dieterle2020PRL}. This allows us to control the ion in a nearly field-free environment for tens of microseconds, sufficiently long to study formation of weakly-bound \Rbmolplus produced via TBR as well as subsequent atom-molecule collisions, and practically limited solely by the level of electric stray field compensation [Fig.~\ref{Fig1}(a)-(c)]. Working in the absence of electric trapping fields offers new means for detecting the inelastic collision dynamics and proves beneficial for the field-sensitive near-threshold molecular ions.

All our experiments start with the production of an elongated BEC of typically $8 \times 10^5$ $^{87}$Rb atoms in the  $|5 S_{1/2}, F=2, m_F=2\rangle$ hyperfine state and held in a magnetic trap with trap frequencies of $(\omega_x,\omega_y,\omega_z)=2 \pi \times (194,16,194) \, \rm{Hz}$. This results in a condensate with Thomas-Fermi radii of $(4.4,53,4.4) \, \mu\rm{m}$ and a peak density of $n_{\rm{at}} = 4.1 \times 10^{14} \, \rm{cm}^{-3}$. We implant a single ionic impurity in the BEC incorporating a precursor Rydberg atom followed by subsequent electric field ionization [Fig.~\ref{Fig1}(a)]. A tailored two-pulse ionization sequence minimizes the initial kinetic energy of the ion to below $k_{\rm{B}}\times 50 \mu$K (for details see Ref. \cite{Dieterle2020PRL}) and strong Rydberg blockade grants the creation of a single impurity \cite{Balewski2013}. Crucially, electric stray fields at the position of the ion are compensated to a level $E\lesssim 300 \, \mu\rm{V/cm}$ in all spatial directions using three pairs of electrodes \cite{Engel2018}. This allows us to let the untrapped ion interact with the host gas atoms for up to tens of microseconds. After a variable evolution time $t$, we apply an electric field pulse of field strength $E_{\mathrm{ex}}= 89 \, \rm{V/cm}$, which interrupts the interaction dynamics, rapidly extracts the ion from the BEC, and guides it towards a microchannel plate detector (MCP). Statistics is gained by repeating the measurement sequence 50 times with the same ensemble of atoms.

\begin{figure}[!ht]
\centering
	\includegraphics[width=\columnwidth]{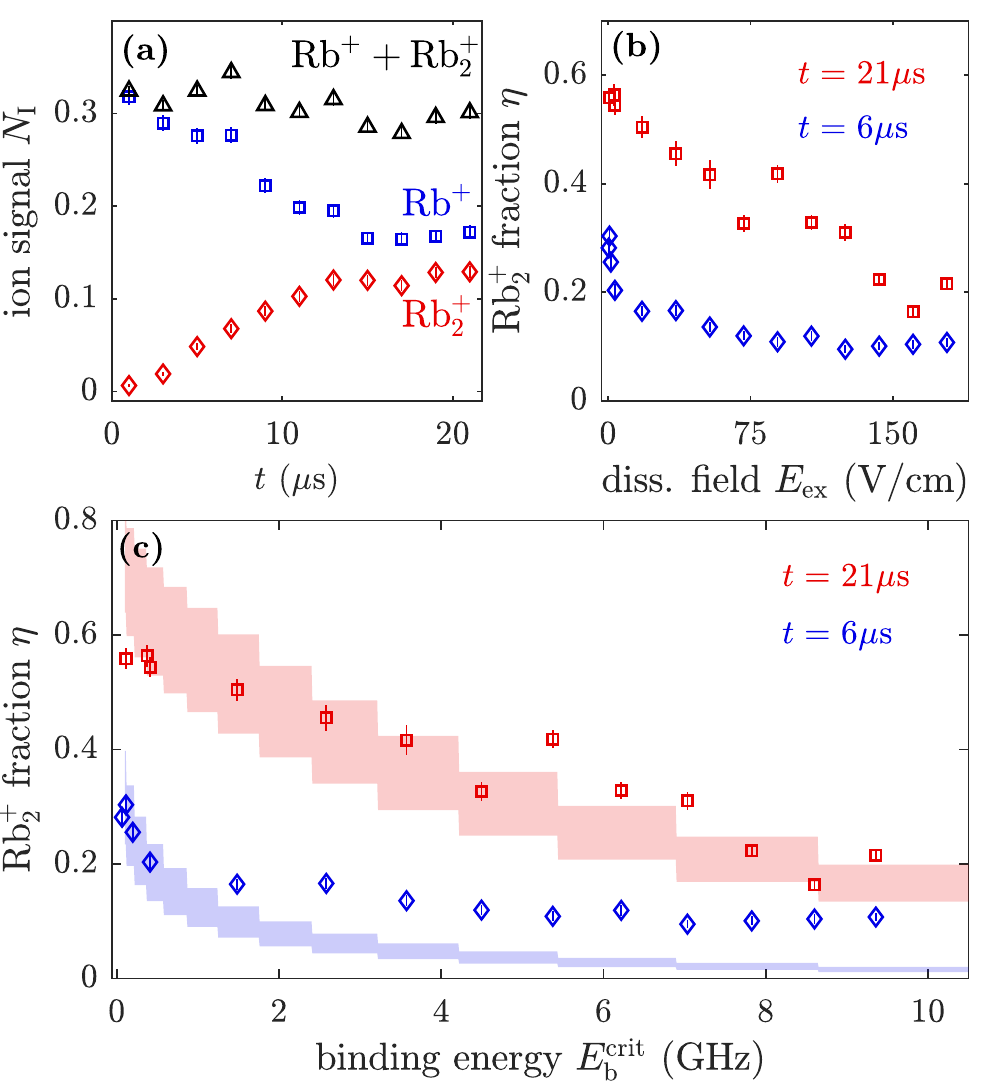}
	\caption{Three-body recombination and molecular ion rovibrational quenching. (a) \Rbplus (blue squares) and \Rbmolplus (red diamonds) signal as a function of evolution time $t$ measured with the \Rbmolplus dissociation field set to $E_{\rm{ex}} = 89 \, \rm{V/cm}$. Black triangles show the sum of the \Rbplus and \Rbmolplus signal. (b) Fraction $\eta$ of detected \Rbmolplus as a function of the dissociation electric field $E_{\rm{ex}}$. Data sets are for $t=\,6\,\mu$s (diamonds) and $t=\,21\,\mu$s (squares).
(c)	$\eta$ as a function of critical molecular binding energy $E_{\rm{b}}^{\rm{crit}}$, which is the threshold value beyond which molecules are dissociated in the electric field $E_{\rm{ex}}$ (see text). The shaded regions are results from numerical simulations of the inelastic collision dynamics.}
\label{Fig2}
\end{figure}

Figure~\ref{Fig1}(e) shows an exemplary histogram of the time of flight to the MCP for $t=21 \, \mu\rm{s}$, exhibiting two distinct peaks associated with the arrival of \Rbplus ions and \Rbmolplus molecular ions \cite{footnote1}. Investigating single experimental realizations, we either detect a \Rbplus ion or a \Rbmolplus molecular ion on the MCP, signifying that molecules appear as a result of a reactive collision process. In the following, we will focus on their formation kinetics in the condensate. The mean \Rbplus and \Rbmolplus ion counts detected on the MCP as a function of $t$, shown in Fig.~\ref{Fig2}(a), provide direct evidence for the formation of molecular ions on a timescale of about $10 \, \mu\rm{s}$. Note that the sum of the two signals (triangles) remains constant over the measurement time. Consequently, for each lost \Rbplus we observe that a \Rbmolplus molecular ion is formed.

Starting from a single low-energy ion immersed in the condensate, \Rbmolplus can only result from a reactive Rb-Rb-\Rbplus TBR collision. In contrast to previous observation of TBR for trapped \Rbplus and \Ba, comprising studies of the energy scaling of the recombination rate coefficient $K_3$ \cite{Haerter2012,Kruekow2016}, our experimental approach provides unique means to explore the reaction products in more detail and moreover measure their subsequent dynamics, complementing very recent observation of intriguing light-assisted reaction kinetics of \BaRb \cite{Mohammadi2020}. Specifically, we exploit the extraction electric field $E_{\rm{ex}}$ to dissociate the molecular ion \cite{Hiskes1961, Blum2016} as illustrated in Fig.~\ref{Fig1}(d). The molecules formed via TBR are only weakly bound and feature large bond lengths. In contrast to the deeply bound rovibrational states, the exchange of electrons between the two atoms is slow compared to the vibrational frequency and the molecule is effectively bound by the ion-atom $C_4$-potential. The resulting asymmetric charge distribution effectively couples to an external field, which allows for strongly perturbing the molecular potential and forcing dissociation, similar to field-ionization of Rydberg states \cite{Bjerre1985}.

Assume the molecular ion is in a certain weakly bound rovibrational state prior to the electric field pulse $E_{\rm{ex}}$. It will then only be detected in the \Rbmolplus channel at the MCP when $E_{\rm{ex}}$ is smaller than the field dissociation threshold of this state [see Fig.~\ref{Fig1}(d)]. Otherwise the molecule is dissociated and the resulting fragment \Rbplus is detected. Consequently, the fraction $\eta$ of detected \Rbmolplus as a function of $E_{\rm{ex}}$ measures the total population of molecular states which are deeper bound than the critical binding energy $E_{\rm{b}}^{\rm{crit}}$ associated with $E_{\rm{ex}}$. Measurements of $\eta$ versus $E_{\rm{ex}}$ are shown in Fig.~\ref{Fig2}(b) for two different values of the evolution time $t$. First, for $t=6\, \mu\rm{s}$ we notice a clear increase in $\eta$ at small values of  $E_{\rm{ex}}$, which points at a tendency that indeed TBR likely produces very weakly bound states. Second, the data set for longer evolution time shows an overall increase in $\eta$ but also indicates redistribution into more deeply bound states. This redistribution is apparent as a qualitative change of the signal. Especially for low dissociation fields, the sharp increase in $\eta$ towards lower dissociation fields in the 6 $\mu$s data turns into a more uniform distribution for longer evolution times.

The redistribution is attributed to secondary inelastic collisions of the molecular ion with neutral atoms leading to quenching into more deeply bound vibrational states. Indeed, the importance of secondary molecule-atom collisions for our experiment can already be seen from a simple estimate. Very similar to elastic ion-atom scattering, collisions of \Rbmolplus with a neutral Rb atom are dictated by a long-range polarization potential, $V(R) = -C_4/(2 R^4)$ \cite{Tomza2019}, giving rates for spiraling-type Langevin collisions $\gamma_{\rm{L}}^{\rm{mol}}=2 \pi n_{\rm{at}} \sqrt{C_4/\mu_{\rm{mol}}}$ \cite{Jachymski2020}. Here, $C_4=318.8$ a.u. is the polarizability for Rb \cite{Holmgren2010}, $R$ is the internuclear separation, and $\mu_{\rm{mol}}$ denotes the reduced mass of the molecule-atom system. Note that for the densities studied in this work, these rates can reach up to hundreds of kHz. The molecule-atom system features additional coupling of molecular states at short range, which inevitably results in inelastic vibrational relaxation for each Langevin collision. The large collision rate $\gamma_{\rm{L}}^{\rm{mol}}$ in our dense gas can then easily yield multiple of such secondary collision events before the untrapped molecular ion exits the condensate, even for the considerable kinetic energy release associated with the \Rbmolplus formation \cite{footnote2}. Note that this is very different for neutral atom systems, for which secondary collisions following TBR-induced loss can be safely neglected.

In a next step, we aim to identify $E_{\rm{ex}}$ more quantitatively with $E_{\rm{b}}^{\rm{crit}}$. To first approximation, one may relate $E_{\rm{ex}}$ with $E_{\rm{b}}^{\rm{crit}}$ using a simple one-dimensional classical barrier model for the dissociation. To this end, we start with the long-range tail of the molecular ion potential in the presence of the external field $E_{\rm{ex}}$ ($q$ denotes the elementary charge)
\begin{equation}
V_{\rm{tot}}(R)=-\frac{C_4}{2\cdot R^4}-\frac{1}{2}\cdot q \cdot E_{\rm{ex}} \cdot R \, ,
\label{eq:ediss}
\end{equation}
which exhibits a field-dependent saddle point [\textit{c.f.} Fig.~\ref{Fig1}(d)]. Dissociation occurs now for all molecular states with binding energy smaller than the height of this potential barrier induced by $E_{\rm{ex}}$. Consequently, one finds for the critical binding energy
\begin{equation}
E_{\rm{b}}^{\rm{crit}} = - \frac{C_4}{2} \left(4\cdot \frac{C_4}{q E_{\rm{ex}}}\right)^{-\frac{4}{5}} - \frac{q E_{\rm{ex}}}{2}\left(4\cdot \frac{C_4}{q E_{\rm{ex}}}\right)^{\frac{1}{5}} \, .
\label{eq:ediss}
\end{equation}

This simple one-dimensional picture, however, only holds for a molecular ion with an internuclear axis aligned along the electric field. In the experiment, we expect the mapping to be altered due to random molecular alignment and molecular rotation. Furthermore, this model assumes a diabatic quench of the initial rovibrational state into a situation where the electric field is instantly present. In the experiment, however, the dissociation field is ramped up within a finite time of typically $\approx 20$ ns. 
We therefore extend our model by performing classical trajectory simulations of the dissociation dynamics, which account for random molecular orientation of the molecule in the laboratory frame and adiabatic state changes upon dissociation. From these simulations, we find a modified relation between $E_{\rm{ex}}$ and $E_{\rm{b}}^{\rm{crit}}$, which follows the functional dependence of Eq.~\ref{eq:ediss} but with an effective value $C_4^{\rm{eff}} = \,10.48 \,\rm{a.u.}$ obtained by fitting Eq.~\ref{eq:ediss} to the numerical results. This modified relation yields the red squares and blue diamonds in Fig.~\ref{Fig2}(c). Consequently, this approach delivers values for $E_{\rm{b}}^{\rm{crit}}$ which only include vibrational dynamics. By adding rotational dynamics to the dissociation model yielding $E_{\rm{b}}^{\rm{crit}}$, we have shown that the qualitative trend of the data does not change.

The data in Fig.~\ref{Fig2}(c) may thus now be compared to a Langevin capture model describing pure vibrational relaxation dynamics, as recently introduced in \cite{Jachymski2020}. Our modeling of the experimental results incorporates both, initial formation of weakly bound molecular ions via TBR followed by secondary two-body collisions of the produced molecule with atoms from the BEC, which result in vibrational relaxation into more deeply bound states. We start out with a TBR collision at a time sampled according to the recombination rate coefficient $K_3$. Subsequently, the product molecular ion undergoes further quenching collisions at random times determined by the Langevin rate $\gamma_{\rm{L}}^{\rm{mol}}$. After each collision the molecule's velocity is increased by the associated kinetic energy release. To keep track of the free motion of the molecular ion through the BEC between successive collisions, we employ a stochastic trajectory approach analogously to our analysis of ionic transport dynamics in a BEC (see Ref. \cite{Dieterle2020PRL}), but now incorporating the two inelastic processes discussed above. Briefly, this approach samples the angular outcome of each collision according to the characteristic isotropic nature of spiraling-type Langevin scattering and further allows us to account for the varying atomic density the molecular ion probes on its path through the BEC.

An integral input to this simulation is a computed probability distribution of product molecular states for the first TBR event as well as branching ratios for the subsequent vibrational relaxation processes. In order to obtain a reasonable estimate for the former, we assume that the three-body collision process leading to molecule formation is sequential, meaning that one atom quenches the two-body ion-atom scattering state, resulting in occupation of multiple bound states with probabilities proportional to the wave function overlaps. This results in a broad distribution of initially occupied bound states, which depends on the typical kinetic energy of the ion-atom pair $E_{\rm{kin}}^{\rm{3b}}$ in our system prior to the recombination and exhibits a maximum corresponding to this value \cite{Kruekow2016}. Note that the more deeply bound molecules resulting from TBR obtain higher initial velocity. The computation of the  branching ratios for the secondary molecule-atom collisions is based on the distorted wave Born approximation as described in Ref.~\cite{Jachymski2020}. In brief, the state-to-state transition probability is proportional to the square of the inelastic elements of the $K$ matrix, which can be approximated as $K_{if} \propto \int dR \phi_i(R)  \phi_f(R) V_{if}(R)$ with $V_{if}$ being the coupling potential between the channels.

Results of these numerical simulations are shown in Fig.~\ref{Fig2}(c) as shaded regions, obtained from typically $10^5$ stochastically sampled trajectories. We find good agreement with the experimental observation on a qualitative level, and even quantitatively the numerical results seem to only slightly underestimate the fraction of more deeply-bound molecular states for the data set taken at $t=6 \, \mu \rm{s}$. Note that the only free parameters entering our model are the value for $K_3$ and the initial kinetic energy of the ion $E_{\rm{kin}}^{\rm{3b}}$, which influences the distribution of the recombination products. Specifically, the results in Fig.~\ref{Fig2}(c) are obtained for $K_3 = 6(2) \times 10^{-25} \, \rm{cm}^6/\rm{s}$ and $E_{\rm{kin}}^{\rm{3b}} = k_{\rm{B}} \times 5(2) \, \rm{mK} $, where the numbers in parentheses reflect the width of the shaded regions. The recombination rate coefficient we obtain here is in fairly good agreement with previous observations using trapped \Rbplus \cite{Haerter2012}, and the value for $E_{\rm{kin}}^{\rm{3b}}$ can be understood from residual stray electric fields accelerating the ion prior to the three-body reaction.
The residual discrepancy at large binding energies between the experimental data at $t=6 \, \mu \rm{s}$ and the numerical simulations hints towards shortcomings in the treatment of the ion-atom-atom recombination process since for short times the distribution of the molecular bound states is essentially determined by the initial three-body recombination event. Within our model one may achieve better agreement with the data for $t=6 \, \mu \rm{s}$ by increasing $E_{\rm{kin}}^{\rm{3b}}$, however, to rather improbable values and also at the expense of accordance with the data for $t=21 \, \mu \rm{s}$.

\begin{figure}[!t]
\centering
	\includegraphics[width=\columnwidth]{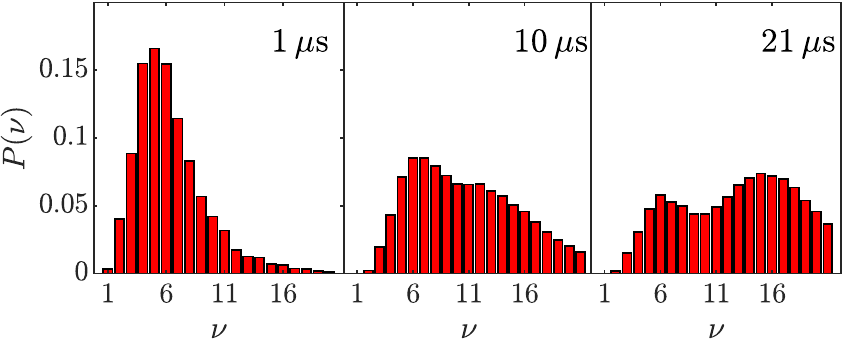}
	\caption{Distribution $P(\nu)$ of the most weakly bound states for three different evolution times obtained from the numerical simulations. The data for $t = 1\, \mu$s reflect the initial distribution of vibrational states occupied as a result of TBR, whereas data for $t = 10\, \mu$s and $t = 21\, \mu$s illustrate redistribution to more deeply bound states via atom-molecule collisions.}
	\label{Fig3}
\end{figure}

Interestingly, our numerical approach delivers the recombination and subsequent vibrational quenching dynamics on the level of individual molecular states. The data in Fig.~\ref{Fig3} shows the obtained distribution $P(\nu)$ of the most weakly bound vibrational states $\nu$ \cite{footnote3}, to which our measurement is particularly sensitive, for three different evolution times and for the values of $K_3$ and $E_{\rm{kin}}^{\rm{3b}}$ given above. In the very early stage of the dynamics $P(\nu)$ reflects the product distribution of the three-body recombination. At later times, the dynamics reveals the two-body vibrational quenching marked by the transfer of population to more deeply-bound molecular states.

In summary, we have investigated the stability of a single ionic impurity embedded in a high density BEC. We have observed inelastic three-body collisions leading to rapid molecular ion formation. We have found evidence for secondary two-body collisions of the weakly bound molecular products and working in an effectively field-free environment allowed us to explore the associated quenching dynamics into more deeply bound rovibrational states by employing a field-dissociation technique in combination with time-of-flight mass spectrometry. Our observations are in good agreement with a model based on Langevin capture dynamics. Evidently, the presented techniques provide valuable features for future experiments with ions in the dense regime of quantum degenerate gases, and may prove beneficial for studies of predicted ionic many-body complexes at lower temperatures \cite{Cote2002,Schurer2017}. Replacing the BEC with a degenerate Fermi gas in future experiments holds exciting perspectives to study the role of quantum correlations of the host gas for the inelastic relaxation dynamics, possibly allowing to increase the lifetime of the ionic impurity. Finally, it seems appealing to elaborate strategies to boost the sensitivity of our field-dissociation probe of the produced weakly bound molecular ions, ideally to a level of resolved quantum states \cite{Haerter2013}, e.g. by exploiting additional microwave spectroscopy \cite{Carrington1996}.

We acknowledge support from Deutsche Forschungsgemeinschaft [Projects No. PF 381/13-1 and No. PF 381/17-1, the latter being part of the SPP 1929 (GiRyd)] and the Carl Zeiss Foundation via IQST. F. M. is indebted to the Baden-W\"urttemberg-Stiftung for the financial support by the Eliteprogramm for Postdocs. K.J. acknowledges support from the Polish National Agency for Academic Exchange (NAWA) via the Polish Returns 2019 programme.

%%%%%%%%%%%%%%-------------------References--------------------%%%%%%%%%%%%%%%%%%


\begin{references}
\bibitem{Tomza2019} M. Tomza, K. Jachymski, R. Gerritsma, A. Negretti, T. Calarco, Z. Idziaszek, and P. S. Julienne, Rev. Mod. Phys. \textbf{91}, 035001 (2019).
\bibitem{Zipkes2010} C. Zipkes, S. Palzer, C. Sias, and M. K\"ohl, Nature \textbf{464}, 388 (2010).
\bibitem{Meir2016} Z. Meir, T. Sikorsky, R. Ben-shlomi, N. Akerman, Y. Dallal, and R. Ozeri, Phys. Rev. Lett. \textbf{117}, 243401 (2016).
\bibitem{Dutta2018} S. Dutta and S.A. Rangwala, Phys. Rev. A \textbf{97}, 041401(R) (2018).
\bibitem{Grier2009} A. T. Grier, M. Cetina, F. Oru\v{c}evi\'{c}, and V. Vuleti\'{c}, Phys. Rev. Lett. \textbf{102}, 223201 (2009).
\bibitem{Feldker2020} T. Feldker, H. F\"urst, H. Hirzler, N. V. Ewald, M. Mazzanti, D. Wiater, M. Tomza, and R. Gerritsma, Nat. Phys. \textbf{16}, 413 (2020).
\bibitem{Schmidt2020} J. Schmidt, P. Weckesser, F. Thielemann, T. Schaetz, and L. Karpa, Phys. Rev. Lett. \textbf{124}, 053402 (2020).
\bibitem{Kleinbach2018} K. S. Kleinbach, F. Engel, T. Dieterle, R. L\"{o}w, T. Pfau, and F. Meinert, Phys. Rev. Lett. \textbf{120}, 193401 (2018).
\bibitem{Casteels2011} W. Casteels, J. Tempere, and J. T. Devreese, J. Low Temp. Phys. \textbf{162}, 266 (2011).
\bibitem{Astrakharchik2020} G. E. Astrakharchik, L. A. Pe\~na Ardila, R. Schmidt, K. Jachymski, and A. Negretti, arXiv:2005.12033 (2020).
\bibitem{Bissbort2013} U. Bissbort, D. Cocks, A. Negretti, Z. Idziaszek, T. Calarco, F. Schmidt-Kaler, W. Hofstetter, and R. Gerritsma, Phys. Rev. Lett. \textbf{111}, 080501 (2013).
\bibitem{Cote2002} R. C\^{o}t\'{e}, V. Kharchenko, and M. D. Lukin, Phys. Rev. Lett. \textbf{89}, 093001 (2002).
\bibitem{Schurer2017} J. M. Schurer, A. Negretti, and P. Schmelcher, Phys. Rev. Lett. \textbf{119}, 063001 (2017).
\bibitem{Esry1999} B.D. Esry, C.H. Greene, and J.P. Burke Jr., Phys. Rev. Lett. \textbf{83}, 1751 (1999).
\bibitem{Burt1997} E. A. Burt, R. W. Ghrist, C. J. Myatt, M. J. Holland, E. A. Cornell, and C. E. Wieman, Phys. Rev. Lett. \textbf{79}, 337 (1997).
\bibitem{Spethmann2012} N. Spethmann, F. Kindermann, S. John, C. Weber, D. Meschede, and A. Widera, Phys. Rev. Lett. \textbf{109}, 235301 (2012).
\bibitem{Haerter2012} A. H\"arter, A. Kr\"ukow, A. Brunner, W. Schnitzler, S. Schmid, and J. Hecker Denschlag, Phys. Rev. Lett. \textbf{109}, 123201 (2012).
\bibitem{Kruekow2016} A. Kr\"ukow, A. Mohammadi, A. H\"arter, J. Hecker Denschlag, J. P\'{e}rez-R\'{i}os, and C. H. Greene, Phys. Rev. Lett. \textbf{116}, 193201 (2016).
\bibitem{Haerter2014} A. H\"arter and J. Hecker Denschlag, Contemp. Phys. \textbf{55}, 33 (2014).
\bibitem{PerezRios2018} J. P\'{e}rez-R\'{i}os and C.H. Greene, Phys. Rev. A \textbf{98}, 062707 (2018).
\bibitem{PerezRios2019} J. P\'{e}rez-R\'{i}os, Phys. Rev. A \textbf{99}, 022707 (2019).
\bibitem{Jachymski2020} K. Jachymski and F. Meinert, Appl. Sci. \textbf{10}, 2371 (2020).
\bibitem{Hudson2016} E. R. Hudson, EPJ Techn. Instrum. \textbf{3}, 8 (2016).
\bibitem{Sullivan2011} S. T. Sullivan, W. G. Rellergert, S. Kotochigova, K. Chen, S. J. Schowalter, and E. R. Hudson, Phys. Chem. Chem. Phys. \textbf{13}, 18859 (2011).
\bibitem{Rellegert2013} W. G. Rellergert, S. T. Sullivan, S. J. Schowalter, S. Kotochigova, K. Chen, and Eric R. Hudson, Nature \textbf{495}, 490 (2013).
\bibitem{Hauser2015} D. Hauser, S. Lee, F. Carelli, S. Spieler, O. Lakhmanskaya, E. S. Endres, S. S. Kumar, F. Gianturco, and R. Wester, Nat. Phys. \textbf{11}, 467 (2015).
\bibitem{Balakrishnan2016} N. Balakrhishnan, J. Chem. Phys. \textbf{145}, 150901 (2016).
\bibitem{Mohammadi2020} A. Mohammadi, A. Kr\"ukow, A. Mahdian, M. Dei{\ss}, J.  P\'{e}rez-R\'{i}os, H. da Silva Jr., M. Raoult, O. Dulieu, and J. Hecker Denschlag, arXiv:2005.09338 (2020).
\bibitem{Dieterle2020PRL} T. Dieterle, M. Berngruber, C. H\"olzl, R. L\"ow, K. Jachymski, T. Pfau, and F. Meinert, Phys. Rev. Lett., submitted (2020).
\bibitem{Balewski2013} J. B. Balewski, A. T. Krupp, A. Gaj, D. Peter, H. P. B\"uchler, R. L\"ow, S. Hofferberth, and T. Pfau, Nature \textbf{502}, 664 (2013).
\bibitem{Engel2018} F. Engel, T. Dieterle, T. Schmid, C. Tomschitz, C. Veit, N. Zuber, R. L\"ow, T. Pfau, and F. Meinert, Phys. Rev. Lett. \textbf{121}, 193401 (2018).
\bibitem{footnote1} The double-peak structure in the molecular ion signal is a consequence of the electrode configuration in the experimental setup. Specifically, a large potential step within an ion lens prior to the MCP leads to dissociation for a fraction of the molecular ions when they pass the lens, which due to their lighter mass appear as the first of the two peaks in the structure.
\bibitem{Hiskes1961} John R. Hiskes, Phys. Rev. \textbf{122}, 1207 (1961).
\bibitem{Blum2016} I. Blum, L. Rigutti, F. Vurpillot, A. Vella, A. Gaillard, and B.  Deconihout, J. Phys. Chem. A \textbf{120}, 20 (2016).
\bibitem{Bjerre1985} N. Bjerre and S. R. Keiding, Phys. Rev. Lett. \textbf{56}, 1459 (1986).
\bibitem{Holmgren2010} W. F. Holmgren, M. C. Revelle, V. P. A. Lonij, and A. D. Cronin, Phys. Rev. A \textbf{81}, 053607 (2010).
\bibitem{footnote2} Note that both the product molecular ion and recoiling atom are not held in the trap after the TBR due to the large energy release. Instead, the products leave the BEC and on their way undergo secondary collisions.
\bibitem{footnote3} We count the quantum number $\nu$ of the vibrational states of \Rbmolplus starting from the dissociation threshold.
\bibitem{Haerter2013} A. H\"arter, A. Kr\"ukow, M. Dei{\ss}, B. Drews, E. Tiemann, and J. Hecker Denschlag, Nat. Phys. \textbf{9}, 512 (2013).
\bibitem{Carrington1996} A. Carrington, Science \textbf{274}, 1327 (1996).

\end{references}
\end{document}